# Modeling PKT at a global level: A machine learning approach

Peetak Mitra
*University of Massachusetts*
Amherst, USA
pmitra@umass.edu

Suhrid Deshmukh, PhD
*Massachusetts Institute of Technology*
Cambridge, USA
suhrid90@gmail.com

*Abstract*—It is well-accepted that the ability to go from one place to another, or mobility, contributes significantly to one's well-being. The need for mobility is universal, but the demand for mobility shows a great variation on a country basis. This particular study looks at what are some of the most important factors on a global level that can help in predicting the passenger-kilometers-travelled or passenger-miles-travelled (PKT/PMT) on a country by country basis. This particular work tries to quantify the impact of some of the key variables like Gross Domestic Product (GDP), population growth, employment rate, number of households, age demographics within the population and macro-economic variables on the total vehicle-based travel within each country. A panel-based regression model is developed to identify the effect of some of the key macroeconomic variables on the countries' PKT growth.

## I. Introduction

Although all the modes of travel are important, this work mainly consists of the study of automobility or mobility as a result of use of cars. Developed countries have a significant vehicular usage and predicting the impact of vehicles or PKT is of critical information for the policymakers, government, land-use planners as well as auto-manufacturers. In most developed countries, travel by a car has become a dominant mode of travel, even as the amount of travel done in non-motorized modes or transit varies by a fair amount. If only the car-based travel is taken into consideration, there are still some significant similarities in the key variables influencing the PMT at a national level by country.

There are quite a few models and studies that look at PKT growth in specific countries. As can be seen from the work done in [1], some of the studies aggregate the data in a bottom up fashion to calculate the nationwide PKT. Some other methods use survey sampling techniques in a similar bottom-up manner as in the previous study to aggregate the PKT calculation [2].

The study of PKT has been done for decades. The literature review here gives an overview of the structural models that have been used to study PKT by various researchers. Most of the studies on PKT are related to land use, highway capacity, the real price of fuel and transit access. The literature on the first three topics is very extensive and exhaustive, so we will limit this review to only meta-analysis models that use summary statistics from primary studies. This is consistent with the method applied in this work where the data was collected from OECD, World-Bank, UN database and the department of transportation and then used to create structural models. A comprehensive review of the PKT models is covered in [4][5][6][7][8].

## II. Dataset

### A. Background

The data for this study was gathered from publicly available sources such as the OECD [3], and the United Nations. We have made the dataset public and it is available to be downloaded from: PKT GitHub repository [https://github.com/peetak/PassengerVKT] and per the knowledge of the authors, is the first of its kind. The data ranges for each parameter vary depending on the parameter itself. While it is advised that one should normalize the data in a regression type problem for better model prediction, it is however not the best approach when it comes to panel data. For the purposes of this study, we have not normalized the data, and the effect of this would be seen in the Results section.

The objective of this work is to quantify the effect of various parameters such as GDP, Population etc., on PKT, over many years. In this dataset, we have gathered data of over 24 countries, about 12 parameters, including Human Development Index (HDI), overall population and population demographics within each country for a particular year. The total number of observations amount to about 26,100. As with any time-series data, the current dataset has the problem of missing values. For such cases, we simply use the *dropna* function within pandas package in Python. More about the code can be found on the PKT GitHub repository.

TABLE I. DATASET OVERVIEW

| Dataset overview | | | |
|---|---|---|---|
| *Number of countries* | *Timeframe* | *Number of parameters* | *Total Observations* |
| 24 | 1970 - 2015 | 12 | 26100 |

## B. Discussion

Before we delve into the model discussion and results, we would like to look at the some of the important features of the dataset and understand why developing a predictive PKT model is hard. The dataset records Passenger Kilometer Traveled, or PKT across 24 countries, including developed as well as developing countries, over a period of 46 years between 1970-2015. The various parameter tracked include total Population, Gross Domestic Product, or GDP, Urbanization, Road Network, Working Age Population, Global Oil Prices and Human Development Index, or HDI, to determine complex effects of these factors into developing a truly robust model for predicting global PKT. To elucidate the complexities in developing a global PKT model we compare data between two groups of countries - developed and developing. Australia and the United States of America, or USA, both of which are considered developed countries and between India and China, countries that are widely considered to be still developing. We make the comparisons between the PKT, GDP and total population in Figures 1-3. The PKT reported in USA, which has the highest GDP in the group, has been eclipsed by India with a higher population and lower GDP since 2005. On the other hand, China with a comparable GDP to USA and higher population does not have a comparable PKT compared to India and USA. It shows the complexities in determining functional relationships between all these factors, that inhibits the development of a truly global PKT model.

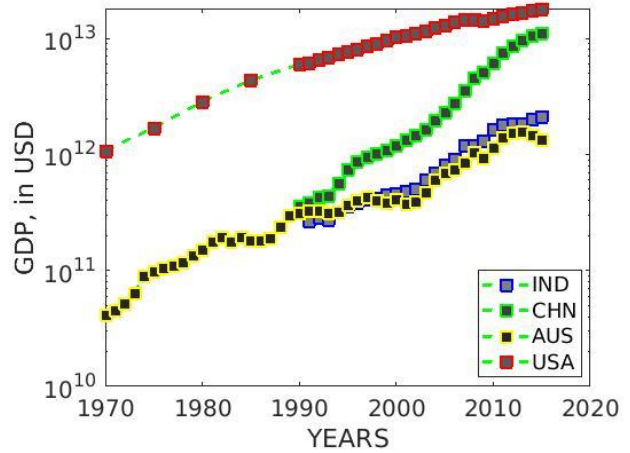

Fig. 2. The GDP evolution shows a cross bunching of data between classes as China catches up with USA, and India-Australia GDP bunched up. IND- India, CHN- China, AUS- Australia, USA- United States of America

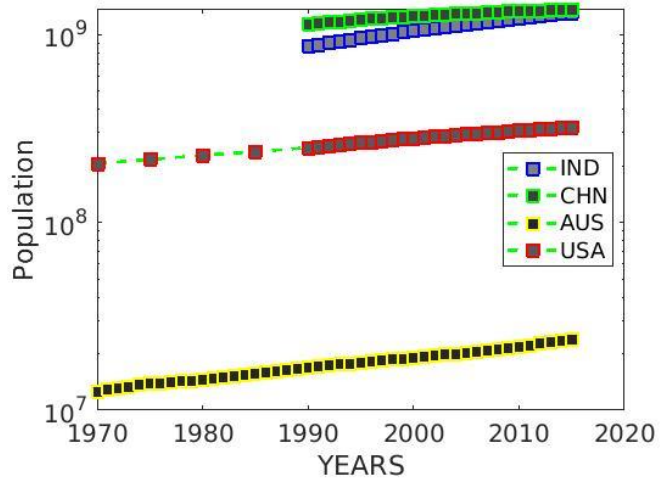

Fig. 3. The Population evolution is on expected lines with countries like India and China leading the way. This however does not co-relate directly to the PKT plots where USA seems to be leading the way. IND- India, CHN- China, AUS- Australia, USA- United States of America

## III. MODEL DESCRIPTION

In panel data, individuals (persons, firms, cities, ...) are observed at several points in time (days, years, before and after treatment, ...). There are two basic models for the analysis of panel data, the fixed effects model and the random effects model. Panel data are most useful when we suspect that the outcome variable depends on explanatory variables which are not observable but correlated with the observed explanatory variables. If such omitted variables are constant over time, panel data estimators allow to consistently estimate the effect of the observed explanatory variables.

One of the main decision points for choosing the model is to decide between fixed and random effects modeling paradigm. In the random effects model, the individual-specific effect is a random variable that is uncorrelated with the explanatory variables. In the fixed effects model, the individual-specific effect is a random variable that is allowed to be correlated with the explanatory variables. Given the kind of data we have, which includes Population, GDP and HDI where there is potentially a high degree correlation within the explanatory variables, we decided to use the Fixed effects model for the statistical analysis.

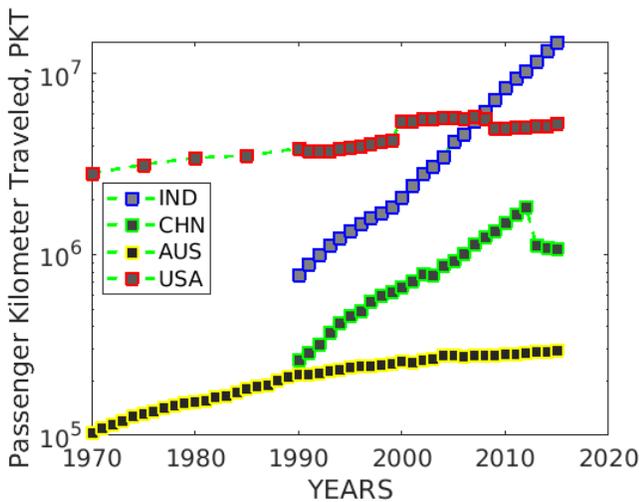

Fig. 1. Passenger Kilometer Traveled evolution for a set of developed and developing countries show no clear trends in terms of grouping of data. IND- India, CHN- China, AUS- Australia, USA- United States of America

Fixed Effects regression is a method that is especially useful in the context of casual inference. While standard regression models provide biased estimates of causal effects if there are unobserved confounders, FE regression is a method that can (if certain assumptions are valid) provide unbiased estimates in this situation. A fixed effects regression is specified on the level of the units and includes group specific constants. Within the fixed effects, we modeled the entity and time effects to create a realistic model framework and the results presented in the next section are based on those modeling assumptions.

The final model comes down to the following form.

$$y_{it} = \alpha_i + \gamma_t + \beta' x_{it} + \epsilon_{it} \quad (1)$$
$\alpha_i$: entity effects
$\gamma_t$: time effects
$x_{it}$: independent variable
$\beta'$: coefficient of independent variable
$\epsilon_{it}$: error term

Having decided on the modeling paradigm, we start off with the hypothesis that PKT is a function of all the input parameters such as below and explore it further using eq (1)

$$PKT = f(GDP, TP, RN, OG, OPC, U, HDI, WA) \quad (2)$$

GDP = Gross Domestic Product
TP = Total Population
RN = Road Network
OG = Oil Prices (Global)
OPC = Oil Prices (Per Capita)
U = Urbanization
HDI = Human Development Index
WA = Working Age

### IV. RESULTS AND DISCUSSIONS

We use the Panel Ordinary Least Squared (OLS) framework in the *linearmodels* package available in Python to model the dataset. We use the formula-based implementation within the *PanelOLS* framework to be consistent with the problem hypothesis (eq. 2). The results of the Panel OLS estimation are shown in Figure 4.

```
                      PanelOLS Estimation Summary
================================================================================
Dep. Variable:                  PKT   R-squared:                        0.7017
Estimator:                 PanelOLS   R-squared (Between):              0.8250
No. Observations:               822   R-squared (Within):               0.7499
Date:              Sat, Jul 13 2019   R-squared (Overall):              0.8211
Time:                      18:19:31   Log-likelihood                   -1.06e+04
Cov. Estimator:          Unadjusted
                                      F-statistic:                      352.79
Entities:                        22   P-value                           0.0000
Avg Obs:                     37.364   Distribution:                   F(5,750)
Min Obs:                     15.000
Max Obs:                     46.000   F-statistic (robust):             472.96
                                      P-value                           0.0000
Time periods:                    46   Distribution:                   F(5,750)
Avg Obs:                     17.870
Min Obs:                     15.000
Max Obs:                     21.000

                             Parameter Estimates
================================================================================
                 Parameter  Std. Err.   T-stat   P-value   Lower CI   Upper CI
--------------------------------------------------------------------------------
GDP             1.388e-07  8.106e-09   17.122    0.0000  1.229e-07  1.547e-07
Urbanization       351.13     761.53   0.4611    0.6449    -1143.9     1846.1
OilPricesGlobal   -37.116     452.97  -0.0819    0.9347    -926.36     852.13
Oil_PerCapitaMWh   3079.6     1380.9   2.2302    0.0260     368.74     5790.5
Population        0.0047     0.0013   3.6922    0.0002     0.0022     0.0072
RoadNetwork      -34.274     87.708  -0.3908    0.6961    -206.46     137.91
================================================================================

F-test for Poolability: 53.468
P-value: 0.0000
Distribution: F(66,750)

Included effects: Entity, Time
```

Fig. 4. PanelOLS estimation summary featuring model performance, as well as parameter importance (based on p-value thresholds)

The model R squared value is about 0.82, which can be considered a good result, since the data is heterogenous and comes from a variety of countries over a long period of time (about 46 years). We model six parameters in this study and per the result, parameters such as GDP, Population and per capita Oil consumption are important metrics for model estimation. This result is on expected lines since the main driver for auto-transport are the population and economics of the country as well as the measure of per capita oil consumption. Other factors such as Urbanization, Global Oil Prices, and Road Network are statistically insignificant for the model. This is an interesting result as one would imagine that Road Network would be an important factor. One needs to investigate this behavior further. Since Global Oil Prices are a metric that might not truly reflect the demand and consumption of oil – an important predictive tool for PKT. On the other hand, the per capita consumption seems like a better metric as it takes into effect additional features such as duties on import of oil for countries such as India, China as well as the strength of the currency, since most of the world's oil is transacted using USD.

The limitations of the current model include the feature set that includes data from 24 countries. There is an opportunity to extend this idea to include many more countries especially from the developing world, including from South east Asia and Africa to make the model a truly robust global level PKT predictor. Including developing countries would mean that factors such as local currency fluctuations, political climate etc. are inadvertently modeled into the system as these affect GDP, HDI, oil prices etc.

### ACKNOWLEDGMENT

The authors would like to acknowledge Google Cloud Platform's Research Credits program for a generous compute grant, that enabled the successful completing of the project.

REFERENCES

[1] A. Hossain and D. Gargett, "Road vehicle-kilometres travelled estimated from state/territory fuel sales," in ATRF 2011 - 34th Australasian Transport Research Forum, 2014.

[2] P. Räty and P. Leviäkangas, "Estimating Vehicle Kilometers of Travel Using PPS Sampling Method," J. Transp. Eng., 2002.

[3] OECD, OECD Factbook 2015/2016 - Economic, environmental and social statistics. 2016.

[4] N. H. T. S. Administration, "Vehicle Survivability and Travel Mileage Schedules," vol. DOT HS 809. 2006.

[5] M. Garrett, S. Bricka, and A. Santos, "National Household Travel Survey," in Encyclopedia of Transportation: Social Science and Policy, 2014.

[6] U.S. Department of Transportation, "Summary of Travel Trends: 2009 National Household Travel Survey," Fed. Highw. Adm., 2011.

[7] M. B. O. Ohlsson, T. S. Rognvaldsson, C. O. Peterson, B. P. W. Soderberg, and H. Pi, "Predicting system loads with artificial neural networks - methods and results from 'the great energy predictor shootout,'" ASHRAE Trans., vol. 100, no. 2, pp. 1063–1074, 1994.

[8] R. Cervero and K. Kockelman, "Travel demand and the 3Ds: Density, diversity, and design," Transp. Res. Part D Transp. Environ., 1997.

[9] R. Ewing and R. Cervero, "Travel and the Built Environment: A Synthesis," Transp. Res. Rec. J. Transp. Res. Board, 2007.

[10] R. Ewing and R. Cervero, "Travel and the built environment," J. Am. Plan. Assoc., 2010.

[11] C. Liu and Q. Shen, "An empirical analysis of the influence of urban form on household travel and energy consumption," Comput. Environ. Urban Syst., 2011.

[12] A. J. Tracy, P. Su, A. W. Sadek, and Q. Wang, "Assessing the impact of the built environment on travel behavior: A case study of Buffalo, New York," Transportation (Amst)., 2011.

[13] D. Salon, M. G. Boarnet, S. Handy, S. Spears, and G. Tal, "How do local actions affect VMT? A critical review of the empirical evidence," Transp. Res. Part D Transp. Environ., 2012.

[14] R. Ewing, "Highway-induced Development Research Results for Metropolitan Areas," Transp. Res. Rec., 2008.

[15] R. Cervero, "Induced travel demand: Research design, empirical evidence, and normative policies," J. Plan. Lit., 2002.

[16] D. J. Graham and S. Glaister, "Road traffic demand elasticity estimates: A review," Transport Reviews. 2004.